\magnification= 1200
\def\rv{{\bf r}}
\def\rp{{\bf r'}}

\def\Ham{{\cal H }}
\def\Graphs{{\cal G }}

\def\half{{ 1\over 2 }}
\def\quart{{1\over 4 }}
\def\3half{{3\over 2 }}

\hsize= 6.0 true in
\vsize= 8.5 true in
\baselineskip 18 pt plus 0.2 pt minus 0.2 pt

\centerline{\bf ROUGHENING INDUCED DECONSTRUCTION}
\centerline{\bf IN (100) FACETS OF CsCl TYPE CRYSTALS}
\bigskip
\centerline{\bf Douglas Davidson and Marcel den Nijs} 
\bigskip
\centerline{\it Department of Physics} 
\centerline{\it University of Washington} 
\centerline{\it Seattle, WA 98195}
\vskip 50 pt 
\noindent
\baselineskip 15 pt plus 0.2 pt minus 0.2 pt

\bigskip
\noindent
The staggered 6-vertex model describes the competition between surface 
roughening and reconstruction in (100) facets of CsCl type crystals.
Its phase diagram does not have the expected generic structure, 
due to the presence of a fully-packed loop-gas line.
We prove that the reconstruction and roughening transitions, cannot  
cross nor merge with this loop-gas line if these degrees of freedom 
interact weakly.
However, our numerical finite size scaling analysis shows that 
the two critical lines merge along the loop-gas line,
with strong-coupling scaling properties.
The  central charge is much larger than 1.5 and 
roughening takes place at a surface roughness
much larger than the conventional universal value.
It seems that additional fluctuations become critical simultaneously.
\bigskip
\bigskip

\centerline{ PACS numbers: 68.35.Rh, 64.60.Fr, 82.65.Dp, 68.35.Bs} 
\vfill
\eject
\baselineskip 18 pt plus 0.2 pt minus 0.2 pt

\centerline{ \bf 1. INTRODUCTION}
\bigskip

In a recent paper, Mazzeo,  Carlon , and van Beijeren [1]
discuss the competition between surface roughening and reconstruction 
in c(2x2) reconstructed (100) facets of CsCl type crystals.
Their numerical finite size scaling (FSS) results
for the staggered 6-vertex model are quite surprising.
The phase diagram lacks a so-called reconstructed rough (RR) phase,
although such a phase is a generic feature in surfaces 
where step excitations do not destroy the reconstruction order.
The phase diagram should have the same structure as for  
missing row (MR) reconstructed simple-cubic (SC) (110) facets [2-4].
The roughening and reconstruction lines should be able to cross.
Instead they only seem to approach each other exponentially close. 
In this paper, we explain why this happens.
The absence of a RR phase in the staggered 6-vertex model 
is accidental, the result of a special symmetry of the interactions 
in this particular model, the presence of a fully-packed loop-gas line.

In section 2, we review the rich history 
of the staggered 6-vertex model.
In section 3,  we describe the topological properties of step and wall 
excitations in c(2x2) reconstructed CsCl(100).
We set up the cell-spin model description for 
this type of competition between surface roughening and reconstruction.
Topological considerations determine whether the 
roughening and  reconstruction lines can cross or only merge
(whether a RR phase is possible or not). 
For example, in MR reconstructed SC(110) facets they can cross,
but in MR reconstructed FCC(110) facets they can only merge [2-4].
We show that in c(2x2) reconstructed CsCl(100) they are allowed to cross.
The competition in this surface is in the same 
universality class as in MR reconstructed SC(110) facets.
However, the RR phase in the staggered 6-vertex model 
is at best narrow. 
We estimate the energies of the two topologically distinct types of steps,
and find that they cost almost the same energy in the 
region of the phase diagram where the surface roughens. 

Carlon {\it et al.} [5,6] point out the existence of a special line
in the phase diagram.
It runs in-between the Ising and roughening lines. 
Along this line the partition function 
is equivalent to the 4-state Potts model on a 
square lattice with negative Boltzmann weights.
They expect this line to be  similar to a so-called disorder line
and that this will explain their numerical results 
(the non-crossing of the Ising and roughening lines).
In section 4 we show that this line is equivalent to 
a fully-packed loop-gas on a square lattice.

In section 5, we prove rigorously that the reconstruction 
line cannot cross the loop-gas line.
Furthermore, we show that the roughening line cannot cross 
the loop-gas line either if the 
roughening and reconstruction degrees of freedom couple weakly.
The weak-coupling hypothesis assumes that 
the reconstruction and roughening degrees of freedom interact weakly,
such that their scaling properties are a superposition.
Earlier studies of models for MR reconstructed SC(110) and FCC(110) 
facets strongly support the weak-coupling hypothesis [2-4].
It should hold for CsCl(100) as well since 
the cell-spin model of section 3 is the same.

This seems to resolve the issue.
The phase diagram found by Mazzeo {\it et al.} [1] is the only one allowed
within weak-coupling theory, but it is an accident.
The special symmetries of the loop-gas line 
cause this particular model to follow a special cut through 
the generic phase diagram.
The roughening and Ising lines approach each other only 
pathologically close, because
entropy cannot be lowered far enough to reach the crossing point
into the RR phase.
In general, the interactions in CsCl type surfaces will be more generic 
and allow the  RR phase.
However, this is not the end of the story.

In section 6 we present our numerical FSS results. 
Mazeo {\it et al.} [1] performed their study before they
discovered the loop-gas line.
Knowledge of the exact location of the line where 
the roughening and reconstruction lines must merge or cross (if they do so)
enhances the accuracy  of the analysis considerably.
We find that the scaling behaviour along the loop-gas line does not obey the
weak-coupling hypothesis.

The question of weak versus strong coupling type competition between 
reconstruction and roughening degrees of freedom is an 
important unresolved issue in the theory of 2D critical phenomena.
It appears not only in surface physics, but also  e.g. in  
coupled Josephson junction arrays in a magnetic field
(the fully frustrated XY model)  [7-13].
The phase diagrams of these problems share as basic feature 
a conventional order-disorder transition line
(such as  an  Ising or 3-state Potts transition),
approaching  a critical (rough) phase.
A critical line described by conformal 
field theory (CFT) with central charge $c<1$
competes with a critical phase  with central charge $c=1$.
The fundamental question is whether new interesting CFT's 
can result from this competition. 
The c-theorem [14,15] implies that such CFT's must have a central 
charge larger than one.
The ones we know are rather simple direct products of 
$c=1$ and $c<1$ theories,
some with  extra symmetries, such as super-symmetry, 
between the two types of degrees of freedom.
This supports the weak-coupling hypothesis.
However, do more interesting types of $c>1$ CFT's really not exist?
Can the coupling between two types of $c\leq1$ degrees of freedom
ever lead to more intricate $c>1$ scaling behaviour?
This question is the driving force behind a large number of numerical studies,
in particular within the context of the fully frustrated XY model
and the competition between surface roughening and reconstruction [1-13].
But it proved to be extremely difficult to answer.

Three types of answers are possible and each has appeared in the literature.
The first one is that the $c<1$ line cannot reach the $c=1$ phase. 
They approach each other only pathologically close. 
This suggests the existence of a no-go theorem of some sort.
Our exact results in section 5 amount to such a no-go theorem, 
but only for the staggered 6-vertex model, 
and only within the weak-coupling hypothesis.

The second possibility is the weak-coupling scenario.
The reconstruction and roughening lines cross or merge, 
but the two types of critical fluctuations interact weakly
and behave like a direct product.
The central charge is equal to the sum
($c=1.5$ in our case, since $c=0.5$ at Ising critical points,
and $c=1$ inside the rough phase). 
This type of behaviour is almost indistinguishable from 
the effective scaling when the two lines approach each other 
only pathologically close (see section 5).
Typically, the numerical data can be interpreted both ways [7-13].
Fortunately, our exact results in section 5
distinguish between the two in  the 6-vertex model.

The third possibility is that the lines cross or merge 
with scaling behaviour different from a simple superposition. 
The central charge is not equal to the sum.
Convincing evidence for strong-coupling would revolutionize CFT at $c>1$.
Numerical evidence for strong-coupling has been presented
in models related to the fully frustrated XY model,
but remains ambiguous [7-13].
We find strong numerical evidence  (in section 6)
that the roughening and reconstruction lines
in the staggered 6-vertex model merge along the loop-gas line, 
with strong-coupling scaling properties.

In section 7, we summarize our results,
discuss related recent results for fully-packed loop-gases on different lattices,
and give a possible explanation for the strong coupling behaviour.

\vfill
\eject

\centerline{ \bf 2. THE STAGGERED 6-VERTEX MODEL}
\bigskip

The (100) facets of CsCl type crystals have a body centered 
type stacking with two kinds of atoms, type A and B.
The appropriate solid-on-solid  description
is a staggered Body-Centered Solid-on-Solid (BCSOS) model, 
equivalent to a staggered six-vertex model. 
Stacks of A atoms occupy the A-sublattice
where the column heights are odd numbers, $h_A=\pm1,\pm3,\pm 5,...$.
The B type atoms occupy the B-sublattice
where the column heights are even, $h_B= 0,\pm2,\pm4,...$.
Nearest neighbour stacks differ in height by only one unit, $dh=\pm 1$.
The surface energy is given by 
$$
\Ham = -{1 \over 4} \sum_{x,y}~ 
\{ E_A ( h_{(x,y)}-h_{(x+1,y+1)})^2 +E_B ( h_{(x+1,y)}-h_{(x,y+1)})^2~\}.
\eqno (1)
$$
The summation runs over only the A-type sublattice sites 
(all even values of $x+y$).

This model has a rich history.
Fig.1 shows its phase diagram.
It is exactly soluble by the Bethe Ansatz along the line $E_A=E_B$,
where it reduces to the so-called F-model [16-18].
We denote the flat phase as $F(n+\half)$, 
because the average surface height is an  half-integer.
A more detailed notation is in terms of the
heights at 4 sublattices,  
$(h_{A,+},h_{A,-};h_{B,+},h_{B,-})=(n+1,n+1;n,n)$.
The $+$ and $-$ indices represent the two  (checkerboard type) sublattices
within each A and B sublattice.
Along $E_A=E_B$, elementary step excitations induce a height change $\pm1$.
They roughen the surface at $z_A=z_B=\half$,
with $z_A=\exp(E_A/k_BT)$ and $z_B=\exp(E_B/k_BT)$ [16-18].

$E_A$ and $E_B$ are not equal in CsCl type surfaces,
since the $A$ and $B$ type atoms interact differently.
Knops [19] realized that this changes the disordering of the $F(n+\half)$ phase.
He used the equivalence of eq.(1) to the Ashkin-Teller model.
Rephrased from the more recent perspective of preroughening (PR) 
transitions and disordered flat (DOF) phases, his results are as follows:
the $F(n+\half)$ contains two types of steps, $S_A$ and $S_B$.
In surfaces with a height $2m+\half$, 
up-steps are of type $S_B$ and down-steps are of type $S_A$
(they  reverse roles in surfaces with a height $2m-\half$).
$-E_A$ is the energy of $S_A$ steps (per unit length), and
$-E_B$ of $S_B$ steps.
The free energy of $S_A$ steps vanishes before that of $S_B$ steps
on the $E_A>E_B$ side of the phase diagram.
This does not cause roughening yet,
since building-up a slope in the surface requires $S_B$ steps as well.
Only the spontaneous symmetry breaking between pairs of surface heights,
$2m+\half$ and  $2m-\half$, is being lifted.
The result is a disordered flat (DOF) phase with lots of $S_A$ type steps 
in the surface, but still flat at large length scales.
In Fig.1, we denote this as the $DOF(2m)$ phase, because 
the average surface height is an even integer, $2m$.

This is an example of a preroughening (PR) transition,
but in a different universality class than the conventional one [2,4].
In both cases the average surface height changes spontaneously by half a unit.
At conventional PR transitions the number of
degenerate equivalent surface heights does not change. 
They all shift by one half unit, and the distance between them remains the same.
In this example however, the distance between degenerate equivalent 
surface heights increases from one to two.
The elementary step height is equal to  $dh=\pm1$ in the F(n+$\half$) phase, 
but equal to $dh=\pm2$ in the $DOF(2m)$ phase.
Therefore, this is a simple Ising line instead of a 
line with continuously varying critical exponents.

This doubling in the basic step height not only changes the nature
of the transition into the DOF phase, it also  delays 
the roughening transition considerably, 
even for small energy differences $E_A-E_B$ [19].
The surface roughness is characterized 
by the amplitude of the height-height correlations as
$$
\left<\left(h_{\rv+\rv_0}-h_{\rv_0}]^2\right)\right> 
\simeq {1\over \pi K_G}  \log(|\rv|).
\eqno(2)
$$
Rough phases become unstable towards discreteness of the surface height
at $K_G=\pi/8$  for step size  $dh=\pm2$ 
compared to $K_G=\pi/2$ for step size $dh=\pm1$,
not until the surface is four times as rough.
In this particular model, these roughening lines lie at the 
side of the phase diagram where  both step energies  are negative
(both $E_A$ and $E_B$ are positive).
Point F in Fig.1 is located at $z_A=z_B=\sqrt(1+\half \sqrt2)= 1.30656$.
The local structure of the phase diagram around point F
is known as a ``critical fan" [20].
Notice that unreconstructed surfaces described by eq.(1)
never roughen (except along $E_A=E_B$).
They follow specific paths through Fig.1,
approximately lines at constant  $E_A/E_B$ 
with both $E_A$ and $E_B$ negative.
Such lines do not enter the critical fan.
The absence of a roughening transition is not a generic feature, however.
Experimental unreconstructed CsCl(100) type surfaces will include
step-step interactions and other aspects that are able to 
move point F towards and beyond $z_A=z_B=1$.
 
In  upper left corner of Fig.1  the surface reconstructs.
All B-type columns are at the same height, $2m$,
and the A-type columns alternate between $2m\pm1$; 
e.g. $(h_{A+},h_{A-},h_{B+},h_{B-}) = (2m+1,2m-1,2m,2m)$.
We call this the $R_A(2m, \theta)$ reconstructed phase. 
The average surface height is an even integer.
The Ising type order parameter $\theta=0,\pi$ 
denotes which of the two A-type sublattices is on top.
The average surface height is the same as in the $DOF(2m)$ phase.
The difference is the
appearance of anti-ferromagnetic type ordering of the A-sublattice heights. 
The competition between the deconstruction of this type of order and
surface roughening is the topic of this paper,
in particular the existence of point $S$ in Fig.1, and the scaling
properties of the critical line beyond this point.

Fig.1 has mirror symmetry with respect to $E_A=E_B$.
The $A$ and $B$ type particles switch roles.
In the lower right corner of Fig.1  the surface height is an odd integer.
The surface  reconstructs into the $R_B(2m+1,\theta)$ phase and
the DOF phase is of type $DOF(2m+1)$. 

\vfill
\eject

\centerline{\bf 3. STEPS AND WALLS IN THE C(2x2) PHASE}
\bigskip
 
Reconstructed surfaces can disorder in several ways:
they can lose their reconstruction first and then roughen;
the can roughen first and only later deconstruct;
or roughening can  induce a simultaneous deconstruction transition. 
The latter happens when the topology of the surface implies
that steps destroy the reconstruction order parameter. 
Fig.2 shows the three topologically distinct line excitations
in the $R_A(2m,\theta)$ phase:
a $(+2,\pi)$ step, a $(0,\pi)$ wall, and a $(-2,0)$ step.
Wall excitations do not change the surface height. 
They cause a switch in which of the two $A$-type sublattices is on top.
Steps of type $(\pm2,0)$ change the surface height by $\pm2$ 
but do not change the Ising order parameter $\theta$;
the same $A$-type sublattice stays on top.
Steps of type $(\pm2,\pi)$ change the surface height by $\pm2$ and switch 
which of the two  $A$-type sublattices is on top. 

It is an illusion to think that $(\pm2,\pi)$ steps destroy the 
reconstruction order.
They preserve a different definition of the reconstruction order. 
It can be expressed in terms of which 
sublattice is on top,  $\theta=0,\pi$, or in terms of 
parity spins, $S_A= \exp[\half i \pi h_A]= \pm 1$.
The $S_A$ spins are ordered anti-ferromagnetically in the 
$R_A(2m, \theta)$ phase.
These two definitions of the Ising order are equivalent 
in flat surfaces but inequivalent in rough surfaces.  
The $(\pm2,\pi)$ steps destroy the sublattice order but preserve the parity order.
The $(\pm2,0)$ steps destroy the parity order but preserve the  sublattice order.
Two types of reconstructed rough (RR) phases are possible:
the surface is rough, 
but such that the $S_A$ order parameter remains non-zero; 
or the surface is rough, 
but such that the $\theta$ order parameter persists.
In diffraction experiments, the roughening transition can be
easily mistaken for a simultaneous deconstruction transition.
The  ``reconstruction diffraction peak"
couples only to one of the two order parameters (typically the parity one) 
and that order might not be the type that persists in the RR phase.

Along paths where either type of step costs less energy than a wall,
the surface roughens first, into the appropriate RR phase,
followed by a Ising type deconstruction transition inside the rough phase.
Along paths where walls cost less than steps, the surface 
deconstructs into the $DOF(2m)$ phase, and only later roughens.

Fig.3 illustrates these different typical behaviours.
It is a schematic phase diagram for the following model [4].
Each site of a square lattice contains  an Ising spin, $\sigma_\rv=\pm1$,
and an height variable , $h_\rv=0,\pm2,\pm4,...$ .
They interact as
$$
\Ham = - \sum_{<\rv,\rp>} 
K_\sigma \sigma_\rv \sigma_\rp +(K_\tau \sigma_\rv \sigma_\rp +K_h) 
\left[ 2- (h_\rv-h_\rp)^2 \right]
\eqno(3)
$$
with $\rv$ and $\rp$ nearest neigbour sites.
Only steps of height $dh=\pm 2$ are allowed.
$\sigma_\rv=\cos(\pi \theta_\rv)= \pm1$ 
represents the sublattice type reconstruction order parameter and
$h_\rv$ the local surface height.
Walls cost $E(0,\pi)= 2(K_\sigma+K_\tau)$.
Steps cost respectively $E(2,0)=  2(K_\tau+K_h)$
and $E(2,\pi)= 2(K_\sigma+K_h)$.
This is a renormalized effective model, 
on a larger length scale than the staggered 6-vertex model.
It must be in the same universality class 
in the local neighbourhood about the reconstructed phase,
assuming we identified correctly all critical fluctuations 
of the 6-vertex model in this part of the phase diagram.
Fig.1 should follow a specific 2D cut through Fig.3. 

The mirror symmetry in Fig.3 with respect to $\Delta= E(2,0)-E(2,\pi)$.
reflects the equivalence between the two definitions of  
the reconstruction order.
Consider the following construction of a typical configuration.
Define a second type of Ising spin $S_\rv=\exp(\half i\pi h_\rv)=\pm1$,
and draw $\sigma$ and $S$ type Bloch walls along the bonds of the lattice.
The $\sigma$ type Bloch walls represent wall excitations in the surface 
and the $S$ type Bloch walls represent $(2,0)$ type steps.
Place one arrow along each $S$ type Bloch wall 
to denote the direction in which the height changes  across the steps
(up from left-to-right while looking in the direction along the arrow).
Sections along the lattice where $S$ and $\sigma$ type Bloch walls 
coincide represent $(2,\pi)$ type steps.
The following change of  variables generates the mirror symmetry in Fig.3.
Define  new Ising spins $\tau=\sigma S$ 
to represent  the $(2,\pi)$ type steps and then 
eliminate the $\sigma$ spins, $(\sigma,S) \to (\tau,S)$. 
This leads us back to eq.(3), but with the $\tau$ spins replacing the $\sigma$ spins,
and with $K_\sigma \leftrightarrow K_\tau$. 
The two  RR phases switched place.

This transformation is reminiscent of super symmetry,
between the fermion (Ising) and boson (height) degrees of freedom.
It is weaker than super symmetry, only a $Z_2$ type invariance [4].
At the same time it is more general, 
an exact symmetry of the lattice model not restricted to $T_c$.
The entire $\Delta=0$ space is invariant,
instead of only the critical line A-B.
We expect that critical fluctuations generate full super symmetry 
at large length scales.
The roughening induced simultanous deconstruction transition
along A-B will then be described by a super symmetric CFT, probably 
one with central charge $c=1.5$ where the roughening and reconstruction
are weakly coupled.

Experimental systems and  microscopic models
follow specific cuts through Fig.3.
For example, the anti-ferromagnetic 
RSOS model describes checkerboard type reconstructed SC(100) facets  [2,4].
Indeed, its phase diagram represents a generic slice out of Fig.3 
with a DOF phase and a RR phase; and
the Ising and roughening degrees
couple weakly with central charge $c=1.5$ [2,4].
One of the exactly soluble generalized RSOS models [21]
moves along the Ising surface in Fig.3 as well
and confirms  weak-coupling behaviour [22].
A third example is the chiral 4-state clock-step model [4] which 
describes MR reconstructed FCC(110) facets.
Topology requires the two types of steps in those surfaces to have 
identical energies.
The non-chiral limit of the 4 state clock-step model
coincides with the $\Delta=0$ plane of eq.(3).
Numerical evidence supports the  
expectation that along A-B the Ising and roughening degrees of 
freedom couple weakly with $c=1.5$. 

The phase diagram of the staggered 6-vertex model 
should to be a generic cut through Fig.3  similar to the RSOS model.
There is no intrinsic topological requirement for $\Delta$ to be zero.
However, Mazzeo {\it et al.} do not find a RR phase.

One possible explanation is that  $\Delta$ is  small
or vanishes  ``accidentally" in the 6-vertex model.
The wall and steps in Fig.2 run diagonally across the surface.
In that direction, the two types of steps cost the same amount of energy
per unit length, 
$E(2,0)=E(2,\pi)= \half \sqrt 2~(2E_A-E_B)$.
A wall cost $E(0,\pi)= \half \sqrt 2~E_A$.
This suggests that $\Delta$ is equal to zero.
However, $\Delta$ is quite large for walls and steps 
running in the horizontal and vertical direction:
$(0,\pi)$ walls cost $E(0,\pi)= E_A$ per unit length, 
$(2,0)$ steps cost $E(2,0)=E_A-E_B$, and
$(2,\pi)$ steps exist only as composite objects 
(a $(2,0)$ step followed by a $(0,\pi)$-wall, i.e., $E(2,\pi)= 2E_A-E_B$).

Walls prefer to run in the diagonal direction, 
but steps switch direction. 
Deep inside the $R(2m,\theta)$ phase
to the left of the line $E_A/E_B\simeq -0.7$
the steps prefer to run in the diagonal direction.
$\Delta$ is small, but this is the part of the phase diagram 
where the walls cost much less energy.
The surface deconstructs into the $DOF(2m)$ phase before it roughens. 
Along  $z_B=0$ the model reduces to the Ising model, and therefore
the deconstruction transition takes place at $z_A=1+\sqrt 2$. 
In zeroth order approximation, the deconstruction line
is located at $z_A=1+\sqrt 2$ for all $E_B$
since the wall energy does not involve $E_B$.

Deep inside the $R(2m,\theta)$ phase
to the right of the line $E_A/E_B\simeq -0.7$
steps prefer the vertical and horizontal directions.
The $(2,0)$ steps are most favourable, and $\Delta$ is large.
However, roughening  cannot take place until
$E_B$ and $E_A$ are of the same order of magnitude. 
Roughening takes place at approximately
$\exp [E(2,0)/k_BT_R]$ $ \simeq1+\sqrt 2 $
(the Ising formula gives reasonable estimates for transition 
temperatures in general).
We can construct two different estimates for the roughening line 
in Fig.1, by assuming the (2,0) steps run vertically or diagonally.
These estimates are quite close to each other.
This means that near the roughening transition  
the $E(2,0)$ steps run almost equally likely in the diagonal 
as in the horizontal or vertical directions.
$\Delta $ must be small near the roughening transition.
$E(2,\pi)$ steps come into play, and 
the RR phase is narrow at best.

\vfill
\eject

\centerline{\bf 4. FULLY-PACKED LOOP-GAS ON A SQUARE LATTICE}
\bigskip

Carlon {\it et al.}[5,6] realized recently  that along the lines 
$z_A+z_B=1$ and $z_A=z_B\pm1$ (the dashed lines in Fig.1)
the staggered 6-vertex model maps onto the 4-state Potts model
and that the $z_A=z_B+1$ line seems to lie in-between the Ising and 
KT roughening lines.
This  mapping has been  known actually for a long time 
but not from this perspective.
For details, we refer to the original source [23].
Carlon and van Beijeren [6] expect that the Potts line will 
turn out to be a type of disorder line, 
and thus will explain that the Ising and  roughening lines cannot meet,
in accordance with their numerical results [1].
The properties of this line are much more intriguing.
The essential observation is that along the Potts line the 6-vertex model 
reduces to a fully-packed (FP) loop-gas.

In the 6-vertex representation of the BCSOS model,
an arrow points along each bond of the lattice, to
denote the height difference between nearest neighbour columns  
$h_A-h_B=\pm1$. Fig.4a shows the 6 allowed vertex states.
In the loop-gas model every bond contains a loop segment.
The loops follow the bonds of the lattice, 
are closed, and  do not intersect.
It is a fully-packed loop-gas.
Fig.4b. shows the 2 possible vertex states, A and B.
The partition function is of the form
$$
Z
=            \sum_{\Graphs} z_A^{N_A}z_B^{N_B} 2^{N_L}
=  z_A^{N_V} \sum_{\Graphs}            x^{N_B} 2^{N_L}, 
\eqno(4)
$$
with the summation over all FP loop-graphs $\Graphs$, 
$x=z_B/z_A$, 
$N_A$ the number of vertices of type A, 
$N_B$ the number of vertices of type B, 
$N_V=N_A+N_B$ the total number of vertices in the lattice, and 
$N_L$ the number of loops.
The fugacity factors of 2 
can be counted by placing arrows on the loops;
clockwise and anti-clockwise arrows.
Loop configurations with such arrows resemble
configurations in the 6-vertex model, but they are not one-to-one related.
The vertex states with anti-parallel arrows, labeled  5 and 6 in Fig.4a,
can be interpreted as both A or B type loop states.
How to deal with this is one of the essential steps
of the mapping of the Potts model onto the 6-vertex model:
each 6-vertex configuration represents the sum over all possible
loop-gas interpretations [23].
Still,  the models are only identical along special lines.
In the 6-vertex model, vertex states 5 and 6 are assigned a Boltzmann factor 
$\omega_5=\omega_6=1$,
(both next nearest neighbour heights are equal), 
while in the loop-gas their weights are the sum over all loop interpretations: 
$\omega_5=\omega_6=z_A+z_B$.
The staggered 6-vertex model reduces to the FP loop-gas, eq.(4), 
only when these are equal, only along the line  $z_A+z_B=1$.

All lines of type $\pm z_A+ \pm z_B=1$ are FP loop gases as well,
due to the fact that vertex states 1 and 2 (and also 3 and 4)  
in Fig.4a appear always in pairs  
(Fig.1 has mirror symmetry with respect to $z_A$ and also $z_B$).
The lines $z_A=z_B+1$ and $z_B=z_A+1$ represent
loop-gases with negative Boltzmann weights
$$
Z
=            \sum_{\Graphs} (-1)^{N_B} z_A^{N_A}z_B^{N_B} 2^{N_L}
=  z_A^{N_V} \sum_{\Graphs} (-1)^{N_B}            x^{N_B} 2^{N_L}. 
\eqno(5)
$$
These two lines are  analytic continuations of each other
with $0 \leq x<1$ along $z_A=z_B+1$ and  $1\leq x <\infty$ along $z_B=z_A+1$;
the minus signs in eq.(5) can be counted  equally well by $N_B$ as $N_A$
since $N_A+N_B=N_V$ is a constant.
FP loop-gases have been a focus of attention recently [24-26].
In particular, the FP loop gas on a honeycomb lattice
resembles eq.(4).
We will discuss possible connections with this recent work 
in section 7.
 
In the loop-gas, the arrows are merely a gimmick to count the loop fugacity. 
They are placed at random on each loop.
Therefore it seems reasonable that any order associated with the 
up-down nature of the steps must be absent along the loop-gas line, 
the reconstruction order as well as the surface flatness.
This is too naive. 
The surface is able to maintain flatness.
Only the reconstruction order is absent.
Notice that the loop gas lines in Fig.1 move through the DOF phases. 
The following visualization is quite useful.
Interpret the $B$ sublattice as  ``patches of red-land" 
and the $A$ sublattice as ``patches of blue-land". 
The loops are the coastlines. 
The two vertex states in Fig.4b represent 
the presence of either a bridge connecting the two patches of blue-land
or the two patches of red-land.
The fugacity factor 2 for each loop
can be interpreted as a random height difference of $dh=\pm1$ between
blue- and red-lands at each coastline;
while looking in the direction along the arrow the land on the left is lower
by one unit than the land on the right.
(The arrows attribute a helicity to each coast line.) 
At $z_A=0$ there exist only red-land bridges.
All red patches are connected and they are all at the same height;
the surface stays flat. 
All blue patches are disconnected and randomly distributed at heights $h\pm1$. 
Consequently, the reconstruction order is absent.
Everywhere along the loop-gas line inside the $DOF(2m)$ phase, 
there exists a large continent of red-land spanning the entire lattice, 
and keeping the surface flat.
Moreover, all blue-lands are finite in size 
(lakes inside the red-land continent).
The red- and blue-lands switch roles inside the  $DOF(2m+1)$ phase.
The only other possibility is that all red and all blue land masses 
are finite in size. 
There the surface is rough and unreconstructed.
The reconstruction order is always absent along the loop-gas line.
In the next section we prove this rigorously.

\vfill
\eject
\centerline{\bf 5. INTERFACE FREE ENERGIES}
\bigskip
Consider the 6-vertex model partition function in a
semi-infinite strip geometry, and
the following boundary conditions (BC's): $h(x+N,y)=\pm h(x,y)+a$,
with $a$ an even integer. 
The lattice forms a cylinder, infinitely long in the $y$-direction
and $N$ lattice sites in circumfence in the $x$-direction.
The free energy per unit strip width for each of these BC's, 
$f(\pm,a)$, can be calculated as the largest eigenvalue of
the transfer matrix.
The free energy differences, $\eta(\pm,a) = f(\pm,a)-f(+,0)$, 
are related to specific step, wall, and other defect free energies.

For example, the 
$h(x+N,y)= h(x,y)+2$ BC forces a step into the $R_A(2m)$ reconstruction.
To see this is, vizualize the 
$R_A(2m)$ ground state as a criss-cross structure
of horizontal and vertical intersecting straight lines. 
Each carries an arrow, pointing  alternately up and down (to the left and right). 
Reversing all arrows interchanges the two degenerate $R_A(2m)$ states.
Reversing the arrows along only one vertical line creates a step. 
This is a $(2,0)$ type step, not a $(2,\pi)$ type step, because
the arrow reversal along this specific line does not affect the 
$\theta$-type order parameter on either side of the step.
The same A-type sublattice stays on top on either side.
By  reversing  the arrow we create a net height difference 
across the surface of $a=\pm2$.
The $h(x+N,y)= h(x,y)+2$ BC matches this structure
for even values of $N$,
and therefore forces a $(2,0)$ step into the surface along the entire cylinder.
$\eta(+,2)$ is equal to the $(2,0)$ type step free energy
everywhere inside the $R_A(2m)$ phase.    

One method to force a wall excitation into the reconstructed phase is
to apply periodic boundary conditions (PBC), $h(x+N,y)=h(x,y)$, 
at odd strip widths $N$.
To see this is, visualize the
$R_A(2m)$ ground state as an array of elementary loops with alternate helicity.
In the  red/blue-land interpretation of the loop-gas the
$R_A(2m)$ ground state is the structure,
in which all blue-lands are disconnected elementary lakes,
and the height changes at the coast lines follow a strict up-down pattern. 
The coast line arrows have alternate helicity.
We run the transfer matrix in the diagonal direction, 
where the reconstructed phase fits only onto the lattice if 
the strip width $N$ is a multiple of 2. 
The  $h(x+N,y)=h(x,y)$ BC frustrates the 
helicity order at odd strip widths.
Therefore  $\eta(+,0)_o$ is equal to the wall free energy. 

The BC $h(x+N,y)=h(x,y)+2$ forces 
a $(2,\pi)$ type step into the reconstructed phase at odd strip widths.
Other types of BC's have similar effects:
twist boundary conditions (TBC) at even values of $N$
create a $(0,\pi)$ wall for  $h(x+N,y)=-h(x,y)$,
and a $(2,\pi)$ type step for $h(x+N,y)=-h(x,y)+2$.

Loop-gas lines in solid-on-solid models signal special properties.
Free energies with certain boundary conditions become ``accidentally" equal,
implying that specific  excitations have identical free energies.
For example, the RSOS model contains a (non fully-packed) loop-gas line, 
which coincides with the exact location of the roughening line. 
Its presence proves the existence of 
the preroughening transition in that model [2].

The partition function of the loop-gas does not change
when we modify the rules for placing the arrows on the loops.
Consider the TBC,  $h(x+N,y)=- h(x,y)$.
The seam is the vertical line across the entire cylinder 
where this boundary condition is being implemented.
(Its location is gauge invariant; moving the seam and deforming it 
does not alter the partition function.) 
The columns on one side of the seam interpret the columns
on the other side as being at height $-h$.
In the arrow representation, this means that the direction of 
the arrow on each loop reverses each time that loop crosses the seam.
There are two types of loops, homotopic and non-homotopic loops.
Non-homotopic loops wrap around the cylinder in such a way that they 
cannot be contracted topologically into a point
(like beads on a necklace).
The requirement that the arrow on each loop reverses each time 
it crosses the seam is incompatible with  non-homotopic loops.
The configurations with PBC, $h(x+N,y)= h(x,y)$, 
are almost identical to those with TBC, $h(x+N,y)=- h(x,y)$.
Each homotopic PBC configuration is matched  by a  TBC configuration.
Moreover, their Boltzmann weights are identical, since
along the loop-gas line the reversal of arrows at the seam
does not affect the Boltzmann factor.
However, all non-homotopic PBC configurations are absent for TBC.
Therefore, the free energies  $f(+,0)$ and $f(-,0)$ 
are identical except for the contribution to $f(+,0)$ of
configurations with non-homotopic loops.
Such configurations are suppressed in the $R_A(2m)$ and  $DOF(2m)$ phase
because all loops  are finite  (all blue-lands are finite sized lakes).
This means that 
$\eta(-,0)= f(+,0,)-f(-,0)$  vanishes  in the thermodynamic limit
exponentially fast with $N$.
The latter is incompatible with reconstructional order, since
$\eta(-,0)$ represents the free energy of a wall excitation in the $R_A(2m)$ phase,
and therefore cannot be equal to zero.
The loop-gas cannot lie inside the $R_A(2m)$ phase. 
  
To make this argument rigorous,
we add the following aspect to the boundary conditions.
Draw a second  seam and associate a phase factor $\phi=\pi/2$ 
($\phi=-\pi/2$) each time a loop crosses this seam with an arrow pointing 
from left to right (right to left).
These phase factors do not affect the Boltzmann factor of homotopic loops, 
(the phases add-up to zero), but they freeze-out non-homotopic loops
(the phases add-up to $\half \pi$ mod($\pi$)).
Define free energies 
$\eta(\pm,a,\phi) = f(\pm,a,\phi)-f(+,0,\phi)$,
with $a=0,\pm2,\pm4$, and $\phi=0,\pi/2$.
From the above discussion it follows that 
along the loop-gas line 
$$
\eta(-,0,0) = \eta(+,0,\half\pi)
\eqno(6)
$$
for all even strip widths $N$.
The twist boundary condition  on the left 
creates a frustration of the Ising order, 
while the periodic type boundary condition
on the right is compatible with Ising order.
This proves rigorously the absence of long-range 
Ising order along the loop-gas line.
The loop-gas line cannot enter the 
$R_A(2m)$ reconstructed phase nor either of the two RR phases.
It cannot cross the Ising line in  Fig.1.

This is a rather weak result.
In particular, it does not imply the absence of a RR phase
because it does not exclude
the possibility that the loop-gas line enters the rough phase.
The implications of eq.(6) become much stronger
within the constraints of the weak-coupling hypothesis.
Next, we summarise the scaling properties along every path 
the loop-gas can follow through Fig.3
and confront each with the loop-gas symmetry eq.(6).
This summary is important also for the numerical analysis in section 6. 
 
(i) Suppose  the loop-gas line enters the deconstructed rough phase.
The central charge is equal to $c=1$, and 
all the above surface free energies  decay at large $N$
as the inverse of the strip width
with amplitudes that are linked to each other as:
$$
\eqalignno{
N \eta_s(+,a,\phi) \simeq & {K_G \over 2} a^2 + {1\over 2K_g} \phi^2 \cr
N \eta_s(-,a,0) \simeq & {\pi\over 4}. 
& (7) \cr
}
$$ 
The inverse of $K_G$, defined in eq.(2), represents  the surface roughness.
The rough phase is described  by the Gaussian fluctuations at large length scales, 
and these relations are exact and easy to derive in the Gaussian model.
$K_G$ must be smaller than $K_G<\pi/8$ since 
the  KT  roughening transition takes place at $K_G=\pi/8$.
This is a factor 4 smaller than the conventional value $K_G=\pi/2$,
because the step excitations create a height difference $dh=\pm2$ instead of
$dh=\pm1$.
The loop-gas identity eq.(6), applied to eq.(7), 
yields the value $K_G=\pi/2$, inconsistent with $K_G<\pi/8$.
The loop-gas line cannot enter the deconstructed rough phase.
(This argument proves also that point P in Fig.1 must coincide with the KT
transition into the $F(n+\half)$ phase.)

(ii) 
Suppose the loop-gas line moves along the Ising surface inside the rough phase,
in particular the phase boundary with the RR phase dominated by $(2,0)$ type steps.
At Ising critical points the central charge is equal to 
$c=0.5$ and the Ising Bloch wall free energy
scales as a power law with universal amplitude
$$ 
N \eta_i \simeq \pi/4.
\eqno(8)
$$
The central charges and universal amplitudes add-up,
as $c = 1.0+ 0.5 =1.5$, when 
the Ising and roughening degrees of freedom interact weakly.
The roughness degrees of freedom behave like in eq.(7), 
with $K_G \leq \pi/8$, 
and the Ising degrees  of freedom like in eq.(8).
This leads to the following FSS amplitudes
$$
\eqalignno{
N\eta(+,2,0)   \simeq & N \eta_s(+,2,0)           \simeq 2K_G           \cr
N\eta(-,0,0)   \simeq & N[\eta_s(-,0,0) +\eta_i]  \simeq \half  \pi     \cr
N\eta(-,2,0)   \simeq & N[\eta_s(-,0,0) +\eta_i]  \simeq \half  \pi     \cr
N\eta(+,0,0)_o \simeq & N \eta_i                  \simeq \quart \pi     \cr
N\eta(+,2,0)_o \simeq & N[\eta_s(+,2,0)_o+\eta_i] \simeq 2K_G +\quart\pi. 
&(9)\cr
}
$$
The loop-gas identity eq.(6), applied to eq.(9), 
yields the value $K_G=\pi/4$, still too large compared to $K_G\leq\pi/8$.
The loop-gas line cannot move along the Ising plane inside the rough phase.
Along the opposite Ising plane,  
the phase boundary with the RR phase dominated by $(2,\pi)$ steps,
the scaling relations are similar, with the two types of steps 
reversing  roles.

(iii) 
Suppose the loop-gas line follows the line segment A-B in Fig.3.
The roughening and Ising degrees of freedom still couple weakly, 
The  central charge remains equal to $c=1.5$,
and the wall free energy scales still as $N\eta_i\simeq \quart\pi$,
but the surface roughenss is constant, $K_G=\pi/8$ [4].
The  FSS amplitudes are similar to those in eq.(9).
The major difference is that 
the amplitudes for the two types of steps 
are identical by symmetry,
$$
N\eta(+2,0)  \simeq  N\eta(+,2,0)_o \simeq 2K_G.
\eqno(10)
$$
The two interface free energies in eq.(6) behave the same as in (ii).
This yields again the value $K_G=\pi/4$, inconsistent with $K_G=\pi/8$.
The loop-gas line cannot move along line segemnt A-B.

(iv) 
Suppose the entire loop-gas line lies inside the DOF phase.
The FSS central charge estimates decay to $c\to 0$.
Both step free energies are non-zero.
The Ising wall type free energy is equal to zero.
However, the convergence becomes extremely slow when the
Ising and roughening lines approach each other closely. 
The asymptotic forms will not be reached, 
and the apparent scaling will be almost indistinguishable from (ii) or (iii).
For  $\Delta\neq0$, 
the effective scaling behaviour will be almost identical to that in eq.(9),
but still be distinguishable;  
the surface roughness cannot exceed the value $K_g = \pi/8$.
For $\Delta=0$,
the effective scaling behaviour will be identical to that for (iii).
This is a fundamental dilemma in studies of this type of phenomena.
The only distinction between scenarios (iii) and (iv) 
is a judgement call on whether the two lines merge or not
in the numerical analysis.
In our case, eq.(6) resolves the issue. 
It excludes (ii) but allows (iv).

In summary, the loop-gas suymmetries exclude all the above scenarios
except the last one.
The only possibilities are:
either the roughening and Ising lines approach each other pathologically close;
or the roughening and reconstruction degrees of freedom
interact strongly and thus circumvent the above arguments.
In any case, the Ising line can never cross the loop-gas line, since
that aspect does not require the weak-coupling hypothesis.

\vfill
\eject

\centerline{\bf 6. Numerical Results}
\bigskip

To distinguish between the scenarios outlined in section 5,
we calculate the exact free energies $f(\pm,a,\phi)$ 
for semi-infinite lattices of width  $2\leq N\leq10$, 
for several boundary conditions as defined in section 4.
We run the transfer matrix in the diagonal direction, 
where the state vector is $2^{2N}$ dimensional.
Such strip widths are in the usual range for transfer matrix calculations. 
They might seem small for readers more familiar with Monte Carlo simulations, 
but realize that our values of 
$f(\pm,a,\phi)$ are accurate to better than 12 decimal places. 
There is no statistical noise, unlike  MC simulations. 
This allows a very detailed finite size scaling (FSS) analysis 
that incorporates the leading corrections to scaling.
It pays to trade system size for the ability to determine 
the leading corrections to scaling, because
at criticality FSS corrections decay only as power laws.
We known the exact location of the line where the reconstruction
and  roughening transitions must merge or cross (if they do),
the FP loop-gas line $z_A=z_B+1$. 
This makes our numerical analysis more accurate than
earlier studies of the same type of phenomena. 

Fig.5 shows the FSS estimates for the central charge 
$c(N)$ along the loop-gas line, $z_A=z_B+1$.
These values follow from the free energy with periodic boundary conditions 
$f(+,0,0)$ and  the conformal field theory scaling relation
$$
f(+,0,0)_N \simeq f_0 + {\pi\over 6 N^2} c 
\eqno(11)
$$
as
$$
c(N) = {6\over\pi} {(N^2-1)^2 \over 4N} [f(+,0,0)_{N-1}-f(+,0,0)_{N+1}]. 
\eqno(12)
$$
Eq.(11) is valid at criticality.
The $c(N)$ approximants of eq.(12),
must converge to zero away from criticality (exponentially fast).
Indeed they do so inside the $DOF(2m)$ phase at small $z_A$.
At criticality $c(N)$ must converge to the  value
characteristic for the universality class of the phase transition.
In Fig.5, we present the raw $c(N)$ data together with
our best FSS estimates for $c$ (the dashed line) and rather conservative
error bars (the shaded area).
At large $z_A$, the  FSS corrections to scaling  become large
and the  FSS analysis of the type $c= c(N)+ A/N^x$ becomes less reliable.
The essential point is that $c$ is certainly larger than $c=1.5$.
This contradicts all weak-coupling scenarios (see section 5).
Maybe $c$ varies continuously along the loop-gas line, 
but it is more likely that $c$ is a constant  between $2\leq c \leq3$ 
and that the variation of $c$ in Fig.5 
reflects crossover scaling behaviour
between that value and  $c=1.5$ at the point 
where the roughening and Ising lines merge.

Fig.6 shows the FSS scaling behaviour of
$N\eta(+,0,0)_o$ and $N\eta(-,0,0)$ along the loop-gas line.
The first one forces an Ising wall into the reconstruction,
the second one an Ising wall and a twist in the surface.
Both free energies vanish in the $DOF(2m)$ phase as expected.
At large $z_A$ they scale with amplitudes that converge 
towards the values $N\eta(+,0,0)_o\simeq\pi/4$ and $N\eta(-,0,0)\simeq\pi/2$
(the dashed lines)
consistent with the weak-coupling scenarios (ii)-(iv).

Fig.7 shows the FSS behaviour of $N\eta(+,2,0)$ and 
$N\eta(+,2,0)_o$ along the loop-gas line. 
$N\eta(+,2,0)$ induces a $(2,0)$ type step in the surface and 
$N\eta(+,2,0)_o$ a $(2,\pi)$ type step.
Both diverge in the $DOF(2m)$ phase as expected
(the step free energies are finite).
At large $z_A$, both decay as power laws. 
These data are inconsistent with scenarios (i)  and  (iii), in which
$N\eta(+,2,0)$ and $N\eta(+,2,0)_o$ should become  equal at large $N$.
The results are inconsistent with scenario (ii) as well, since
$N\eta(+,2,0)$ and $N\eta(+,2,0)_o$ should differ by $\pi/4$ when the 
loop-gas line moves along the Ising plane inside the rough phase (see eq.(9)).
To demonstrate this, we plot in Fig.7 
$N[\eta(+,2,0)_o -\eta(+,0,0)_o]$ instead of  $N\eta(+,2,0)_o$ itself.
The two sets of curves should fall on top of each other
and converge towards a continuously varying  roughness parameter $2 K_G$.
Instead, they differ by a factor close to 2.
Most importantly, the data is inconsistent with scenario (iv),
in which the roughening and Ising lines only approach each other 
asymptotically close.
We should find effective scaling behaviour of the form
$N\eta(+,2,0) \simeq N[\eta(+,2,0)_o -\eta(+,0,0)_o]\simeq 2 K_G$
with  an effective  roughness $K_G\geq\pi/8$.
Not only do the two amplitudes differ by a factor 2,
both sink well below $2K_G=\pi/4$.
The surface roughness becomes too large by a factor of about two.

Fig.8 shows an example, of the FSS behaviour of $N\eta(+,2,0)$ 
along the cut $z_A+z_B=11$ through the loop-gas line.
There exists no RR phase on either side of the loop-gas line.
In the reconstructed phase, $N\eta(+,2,0)$ must diverge and  
in the rough and RR phases converge to $2K_G~$.  
At the  KT roughening transition, $2K_G$ must be equal to $2K_G=\pi/4$.
The conventional method to determine the roughening temperature is
to extrapolate the points where  $N\eta(+,2,0)=\pi/4$ as function of $N$.
In Fig.8, these points lie at the reconstructed side of the loop-gas line.
They converge towards the loop-gas line
at such a rate that power law fits actually overshoot the loop-gas line.
This might lead to the  conclusion that the roughening line does not 
cross nor merge with the loop-gas line.
On the other hand, along the loop-gas line itself, the amplitude converges 
very well to a value much smaller than $\pi/4$.
This behaviour is similar to what happens at a conventional 
KT type roughening transition if one tries to estimate $T_c$ by
extrapolating the points where the FSS amplitude 
is larger than the true critical value,
$N\eta(+,2,0)=\pi/4+a$  with $a>0$.  
The FSS behaviour in Fig.9 strongly supports the 
absence of a RR phase. 
$N\eta(+,2,0)$ diverges everywhere
on the reconstruction side of the loop-gas line.
Roughening seem to take place exactly at the loop-gas line but 
surprisingly at a surface roughness well above the universal KT value.

Fig.9 shows two types of estimates for the
critical point where the roughening and Ising lines merge:
the $N\eta(-,0,0)$ crossing points from Fig.6, and 
the $N\eta(+,2,0)=\pi/4$ points from Fig.7.
The existence of crossing points in $N\eta(-,0,0)$ and $N\eta(+,0,0)_o$
(see Fig.6) is quite significant.
In the more global context of Fig.1, these
crossing points converge to the Ising type reconstruction lines.
(This is the conventional method to locate such critical lines.)
Their presence along the loop-gas line implies that all FSS estimates 
for the Ising transition cross the loop-gas line.
The existence of points along the loop-gas line 
where $N\eta(2,0,0)=\pi/4$ (see Fig.7) is equally significant.
These points converge to the KT roughening lines in Fig.1. 
(This is the conventional method to determine such roughening lines.)
Their presence along the loop-gas line 
implies that all FSS estimates for the roughening transition 
cross the loop-gas line as well
(see also the discussion about  Fig.8).
The Ising and roughening lines cross the loop-gas line
from opposite directions.
All finite $N$ estimates for the Ising and roughening lines 
cross each other.

Both curves in Fig.9 must converge to $E_c\to \infty$ 
if the roughening and Ising lines only approach each other asymptotically
close.
This is very unlikely, although both curves are convex
(the corrections to scaling behave as $E_A(N)=E_c+A/N^x$  with  an exponent $x<1$).
A conservative estimate puts the critical point somewhere 
between $1.4<E_c<1.6$. 

These numerical results contradict all weak-coupling scenarios, 
in particular the one  where the roughening and Ising lines 
only approach each other pathologically close,
the only one allowed by the loop-gas symmetries of section 5.
The most damaging evidence is that  
$N\eta(+,2,0)$ becomes too small by a factor 2.
It seems too far fetched that this amplitude can rebound 
all the way back to $\pi/4$ at very large $N$.
Moreover,  the central charge is significantly larger than 
the weak-coupling value $c=1.5$.
Finally, the FSS estimates for the Ising and roughening line
cross each other for all finite $N$, and lead to an estimate
$E_c\simeq 1.5\pm0.1$ for the point along the loop-gas line where
they merge.

\vfill
\eject
\centerline{\bf 7. CONCLUSIONS.}
\bigskip

In this paper, we study the phase diagram of the staggered 6-vertex model
from the perspective of the competition between surface roughening
and reconstruction, and also in the context of  unresolved issues
about the scaling properties of QFT with central charge $c>1$.
In section 3, we review  the weak-coupling hypothesis as
encountered in previous studies for this type of interplay. 
In particular we show that a reconstructed rough (RR) phase must be
present in typical c(2x2) CsCl(100) type surfaces.
The staggered 6-vertex model includes a fully-packed (FP) loop-gas line
(section 4). 
Its special symmetries explain the absence of the RR phase
in this particular model.
In section 5, we prove that the Ising type reconstruction line
cannot cross the loop-gas line. 
Moreover, we demonstrate that within the context of 
the weak-coupling hypothesis the 
roughening line  cannot cross the loop-gas line either.
This would explain the results by Mazzeo {\it et al.}  and 
put them in agreement with the generic phase diagram Fig.3.
However, our numerical results contradict all weak-coupling scenarios.
The Ising and roughening lines merge along the loop-gas line.  
The central charge is large, at least equal to $c=2$ (see Fig.5), and the 
surface roughness increases to a value about twice as large
as the universal KT value (see Fig.7).

Eqs.(4) and (5)  represent FP loop-gases on a square lattice.
The FP loop-gas model on a honeycomb (HC) lattice  
and a four-coloring problem on the square lattice are related to this.
Those models have large central charges as well, 
respectively $c=2$ and  $c=3$ [24-26].
Unfortunately they seem more closely related to eq.(4) than eq.(5).
The partition function of 
the  FP loop-gas on a HC lattice is similar to eq.(4), 
with three types of ``bridge-energies" instead of two,
$z_\alpha$ with $\alpha$=A,B,C (the plaquettes of the HC lattice 
form three sublattices instead of two).
Point $P$ in our phase diagram is a critical point with central charge $c=1$. 
It is the meeting point of two DOF phases and also the
point where the $E_A=E_B$ surface roughens (see Fig.1).
The corresponding  point in the FP loop-gas on the HC lattice, $z_A=z_B=z_C$,  
is a critical point with central charge $c=2$ [24-26].
It is the meeting point of three DOF phases, and of
three critical lines with central charge $c=1$
(the phase boundaries between pairs of DOF phases) [27].
The scaling properties along $z_A=z_B\pm 1$ in Fig.1 
appear to be  more complex than this.
This is not surprising since eq.(5) includes negative Boltzmann weights.

We close with some speculations about the origin of strong-coupling 
scaling along the loop-gas lines  $z_A=z_B\pm 1$.
It might represent a novel type of conformal field theory with $c>1$.  
More likely, it represents a conventional CFT, but one 
in which more degrees of freedom become critical
than in  eq.(3) and Fig.3.
The obvious candidate is the Ising degree of freedom
of the $R(2m+1,\theta)$ phase located on the opposite side
of the $E_A=E_B$ line in Fig.1.
The two reconstructed  phases do not seem to meet in
Fig.1, but they actually do so via the back-door.
They meet at point $x=1$ in eq.(5),
since  $z_A=z_B+1$ and $z_B=z_A+1$ are each other's analytic continuations.
A 4-state clock model coupled to a SOS model describes
both types of reconstructions simultaneously.
The $\sigma=\pm 1$ Ising spins in eq.(3) denote which of the
two A-type sublattices is on top in the $R(2m,\theta)$ phase.
Their generalizations are 4-state clock variables, $\theta=0,\pm \half \pi, \pi$,
that point in the direction of the polarization of the arrows 
in vertex states 1-4 in Fig.4a.
They denote which of the four sublattices is no top:
one of the two A-type sublattices, $\theta=0,\pi$;
or one of the two B-type sublattices $\theta=\pm\half\pi$.
Such a model has ample room for conventional CFT's with central charge $c\geq 2$.
The loop-gas symmetries will enforce a non-generic path through its phase diagram.
Point $x=1$ in eq.(5), where the two reconstructions meet, 
is almost certainly a critical point with central charge $c=2$.
Fig.3  applies when $x=1$ is an isolated critical point.
In that case, the Ising and roughening lines cannot meet until $x=1$
(our exact results of section 5).
Instead, our numerical analysis shows that $x=1$ is not an isolated point.
The next step will be an analytic calculation, 
to determine the scaling properties at point $x=1$.
The critical dimension of the crossover operator in the loop-gas direction
must be irrelevant or marginal for the
loop-gas to remain critical until point $S$ in Fig.1.
This line segment is probably  some some sort of Baxter-line coupled to 
roughening, since  the universal amplitudes of both step free 
energies in Fig.7 vary continuously.

It is a pleasure to thank Enrico Carlon,
Georgio Mazzeo, and  Henk van Beijeren, for many discussions.
This research is supported by NSF grant DMR-9205125.

\vfill
\eject

\centerline{ \bf REFERENCES}
\bigskip
\item{ 1. } 
G.~Mazzeo, E.~Carlon, and H.~van Beijeren, Phys.Rev.Lett.~{\bf 74},1391 (1995).
\item{ 2. } 
M.~den Nijs and K.~Rommelse Phys.Rev.~B {\bf 40}, 4709 (1989);
\item{    } 
M.~den Nijs, Phys.Rev.Lett.~{\bf 64}, 435 (1990).
\item{ 3. } 
For a review, see  M.~den Nijs, chapter 4 in
{\it Phase Transitions and Adsorbate Restructuring at Metal Surfaces}, 
of {\it The Chemical Physics of Solid Surfaces}, Vol.7, 
D. King editor (Elsevier, 1994).
\item{ 4. } 
M.~den Nijs, Phys.Rev.~B {\bf 46}, 10386 (1992).
\item{ 5. } 
Enrico Carlon Ph.D.~thesis, University of Utrecht, The Netherlands (1996).
\item{ 6. } 
Henk van Beijeren, Enrico Carlon, and Georgio Mazzeo, private communication.  
\item{ 7. } 
T.C.~Halsey, J.Phys.~C {\bf 18}, 2437 (1985).
\item{ 8. } 
J.M.~Thijssen and H.J.F.~Knops, Phys.Rev.~B {\bf 37}, 7738 (1988);
\item{    } 
and Phys.Rev.~B {\bf 42}, 2438 (1990).
\item{ 9. } 
E.~Granato and M.P.~Nightingale, Phys.Rev.~B {\bf 48}, 7438 (1993).
\item{10. } 
Y.M.M.~Knops, B.~Nienhuis, H.J.F.~Knops, and H.W.J.~Blote, 
Phys.Rev.~B {\bf 50}, 1061 (1994).
\item{11. }   
M.P.~Nightingale, E.~Granato, and J.M.~Kosterlitz,
Phys.Rev.~B {\bf 52}, 7402 (1995).
\item{12. } 
P. Olsson, Phys.Rev.Lett.~{\bf 75}, 2758 (1995).
\item{13. }
H-J.~Xu and B.W.~Southern, J.~Phys.~A {\bf 29}, L133 (1996).
\item{14. }
A.B.~Zamolodchikov JETP Lett.~{\bf 43}, 730 (1986). 
\item{15. }
I.~Affleck Phys.~Rev.Lett.~{\bf 56}, 746 (1986).
\item{16. }
H.~van Beijeren, Phys.Rev.Lett.~{\bf 38}, 993, (1977).
\item{17. }
H.~van Beijeren and I.~Nolden, in {\it Strucures and dynamics of Surfaces},
edited by W.~Schommers and P.~von Blanckenhagen  Vol.~2 
(Springer, Berlin 1987).
\item{18. }
E.H.~Lieb and F.Y.~Wu, 
in {\it Phase Transitions and Critical Phenomena}
eds. C.~Domb and M.S.~Green (Academic, London, 1972).
\item{19. }
H.J.F.~Knops, J.~Phys.~A {\bf 8}, 1508 (1975).
\item{20. } 
M.~Kohmoto, M~.den Nijs, and L.P.~Kadanoff, Phys.Rev.~B {\bf 24}, 5229  (1981).
\item{21. } 
A.B.~Zamolodchikov and V.A.~Fateev, Sov.Nucl.Phys.~{\bf 32}, 298 (1980).
\item{22. } 
T.T.~Truong and M.~den Nijs, J.Phys.~A {\bf 19}, L645 (1986).
\item{23. } 
R.J.~Baxter, S.B.~Kelland , and F.Y.~Wu, J.Phys.~A {\bf 9},397 (1976).
\item{24. } 
H.W.J.~Blote and B.~Nienhuis, Phys.Rev.Lett.~{\bf 72}, 1372 (1994);
\item{    } 
and Phys.Rev.Lett.~{\bf 73}, 2787 (1994).
\item{25. } 
J.~Kondev and C.L.~Henley, Phys.Rev.Lett.~{\bf 73}, 2786 (1994);
\item{    } 
and Phys.Rev.~B {\bf 52}, 6628 (1995). 
\item{26. } 
J.~Kondev, J.~de Greer, and B.~Nienhuis, J.Phys.~A,  in press (1996).
\item{27. }
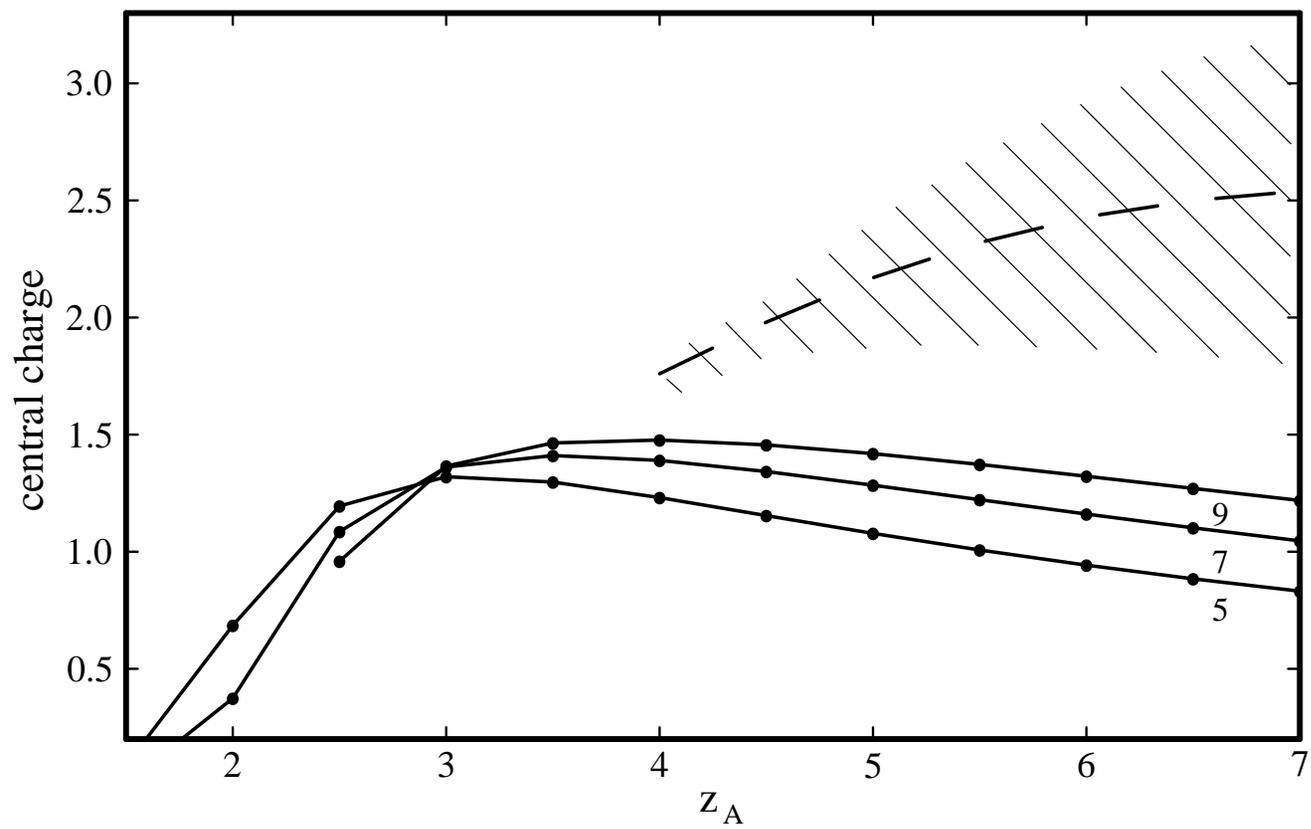
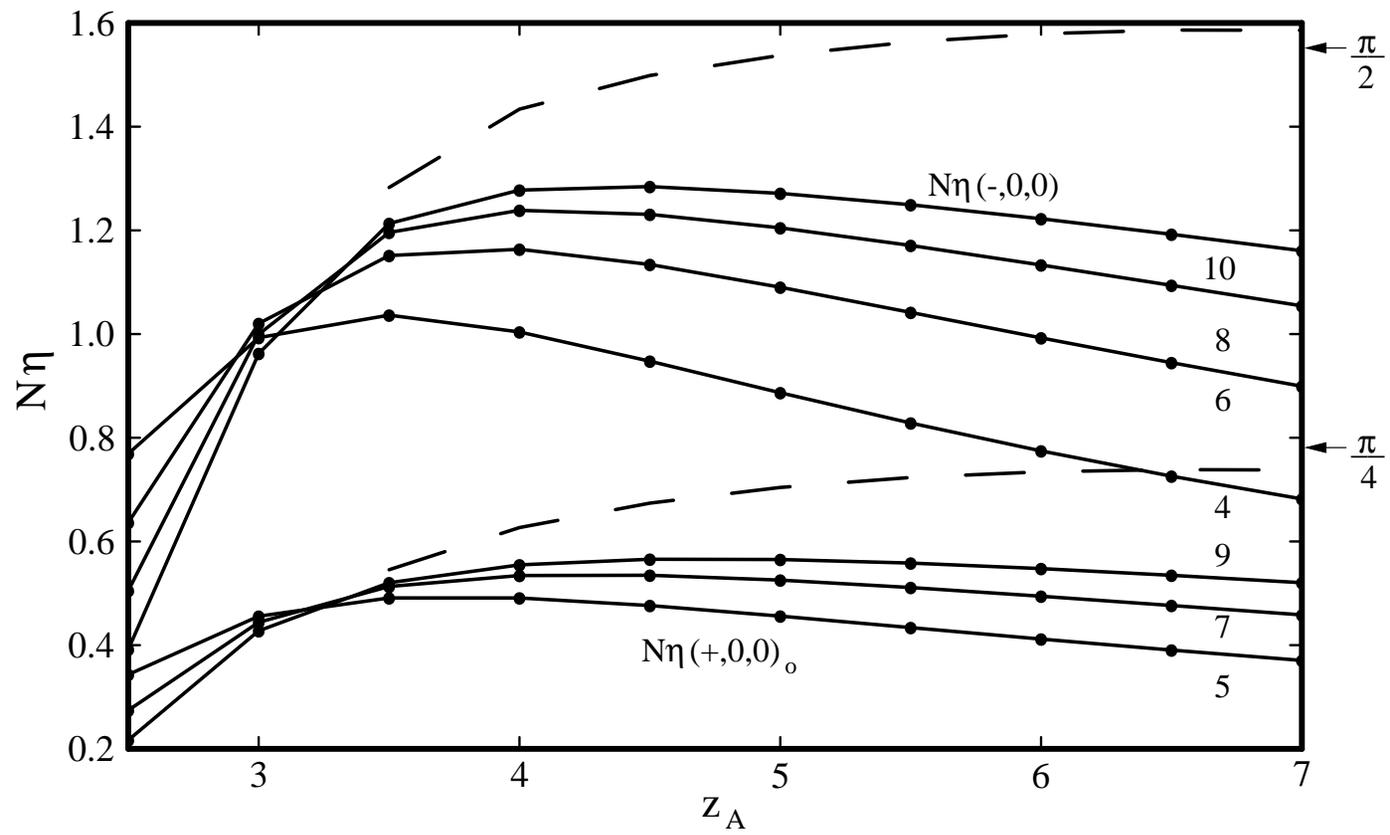
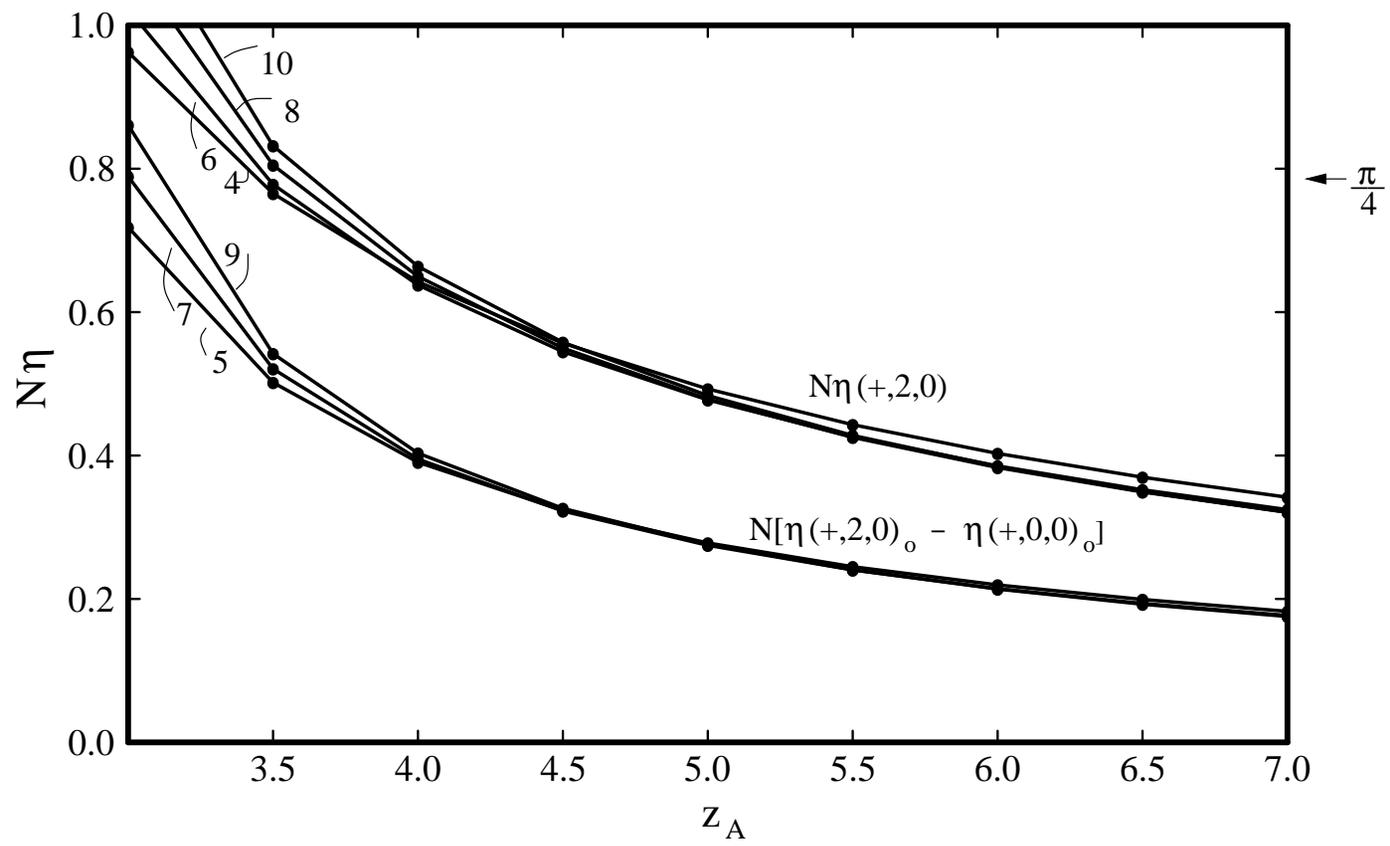
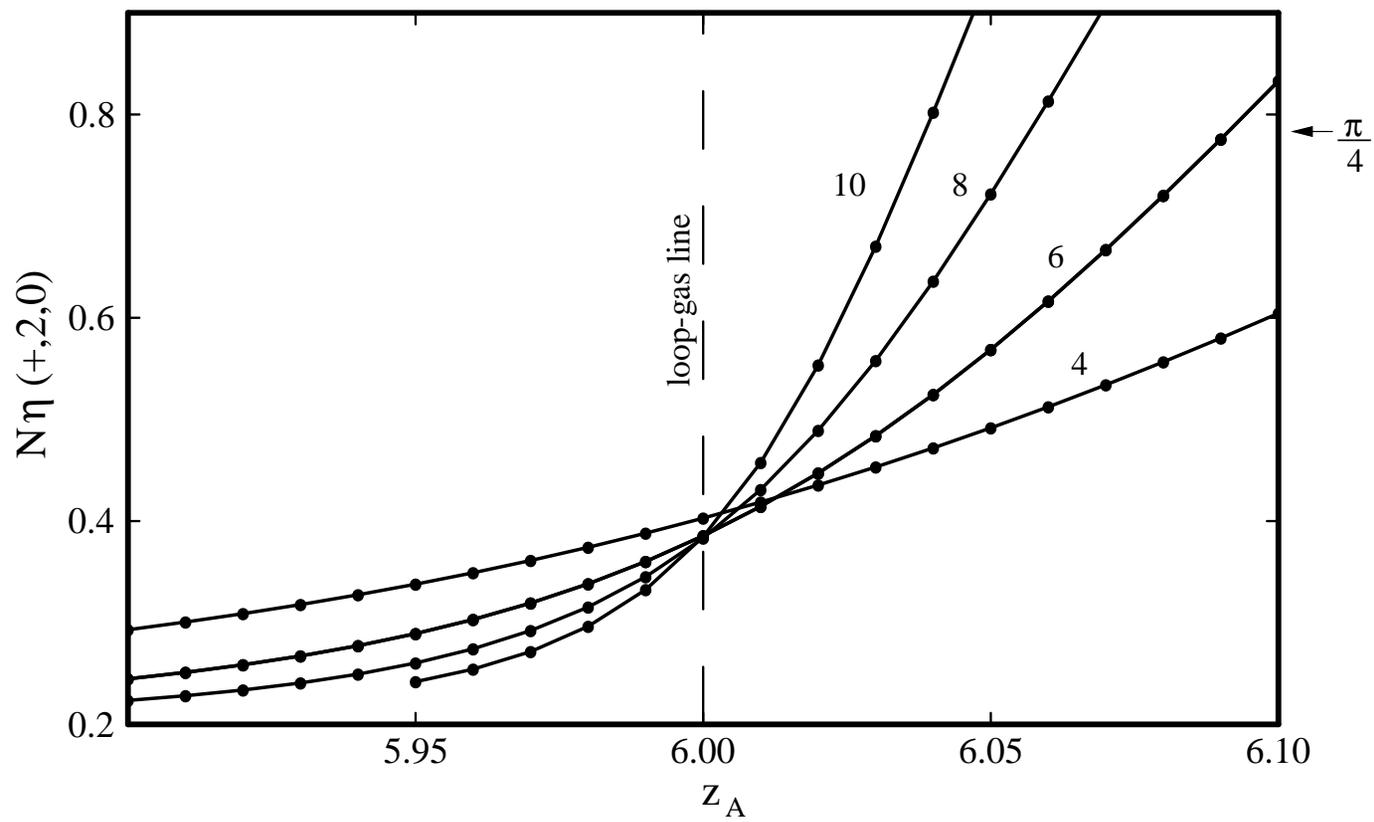
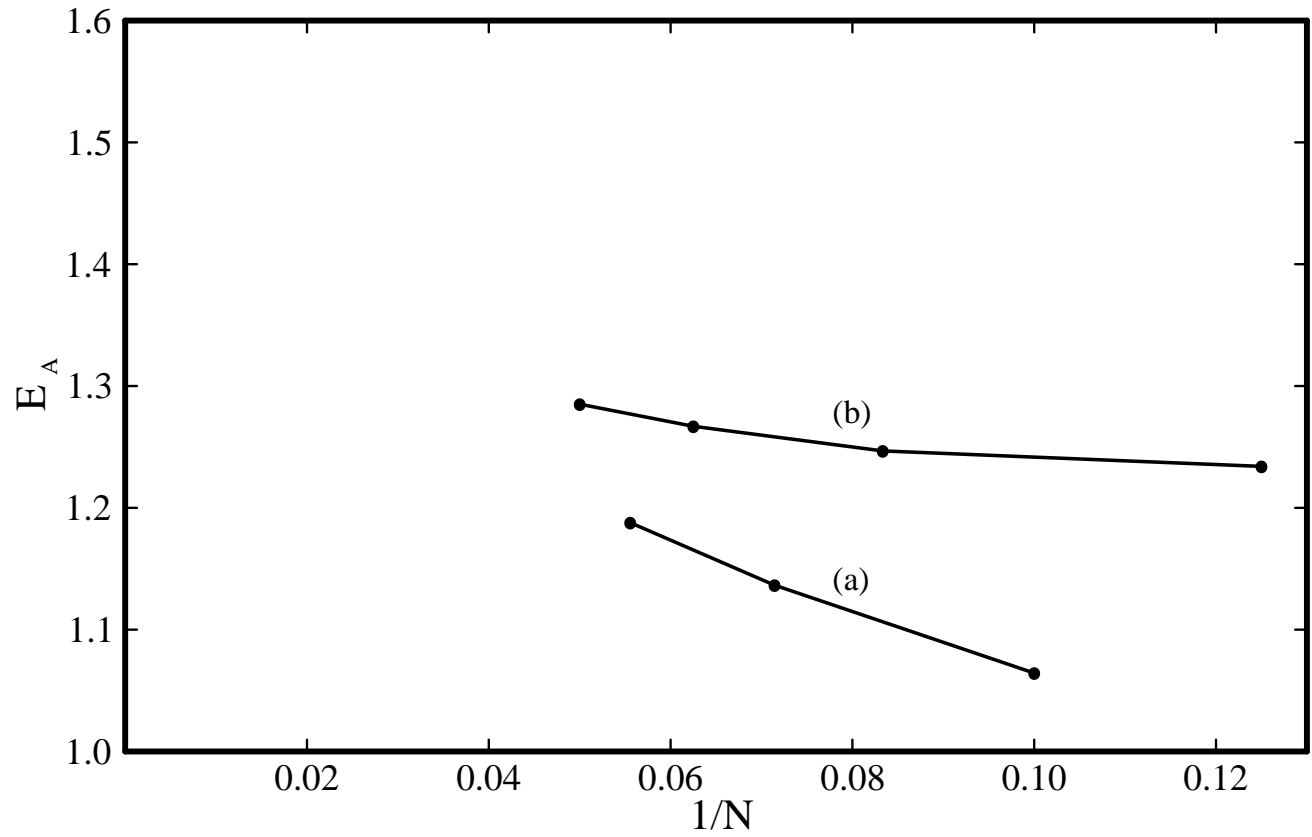
D.~Davidson and M.~den Nijs, unpublished (1996). 
\vfill
\eject

\centerline{ \bf FIGURE CAPTIONS}
\bigskip
\item{ 1.} 
Phase diagram of the staggered 6-vertex model defined in eq.(1).
At point S the Ising reconstruction and KT roughening lines merge
into a single transition along the loop-gas line (dashed).
\item{ 2.} 
Excitations inside the  $R_A(2m,\theta)$ reconstructed phase: a 
$(2,0)$ step, $(0,\pi)$ wall, and $(2,\pi)$ step
running in the diagonal direction.  
\item{ 3.}
Generic phase diagram for the competition between surface roughening and 
Ising type reconstruction, see eq.(3), 
with 
$\Delta$ the energy difference between $(2,0)$ and $(2,\pi)$ type steps, 
and
$R$ the  difference between the energy of a  $(0,\pi)$ wall 
and the average step energy.
\item{ 4.}
The six vertex states of the 6-vertex model (a),
and the two vertex states of the loop gas model (b), 
with their Boltzmann weights.
\item{ 5.}
FSS estimates for the central charge along the loop gas line
for strip widths $N\leq10$.
The shaded area represents a conservative estimate for the
uncertainty in the extrapolated values (the dashed line).
\item{ 6.}
$N\eta(+,0,0)_o$ (free energy of a $(0,\pi)$ wall) and $N\eta(-,0,0)$ 
(free energy of a $(0,\pi)$ wall with a twist in the surface)
along the loop gas line for strip widths $N\leq 10$.
\item{ 7.}
$N\eta(+,2,0)$ (free energy of a $(2,0)$ step) 
and $N\eta(+,2,0)_o-N\eta(+,2,0)_o$
(free energy of a $(2,\pi)$ step minus that of a $(0,\pi)$ wall).
along the loop gas line for strip widths $N\leq 10$.
\item{ 8.}
The step free energy $N\eta(+,2,0)$ along the line $z_A+z_B=11$
for strip widths $N\leq 10$.
\item{ 9.}
FSS estimates for the location of the 
critical point $\exp(-E_c)=z_A=z_B+1$ along the loop gas line from:  
(a) the $N\eta(-,0,0)$ crossing points in Fig.6, and 
(b) the $N\eta(+,2,0)=\pi/4$ points in Fig.7.

\vfill
\eject
\end